# Head/tail Breaks: A New Classification Scheme for Data with a Heavy-tailed Distribution

Bin Jiang
Department of Technology and Built Environment, Division of Geomatics
University of Gävle, SE-801 76 Gävle, Sweden
Email: bin.jiang@hig.se

**Abstract**
This paper introduces a new classification scheme – head/tail breaks – in order to find groupings or hierarchy for data with a heavy-tailed distribution. The heavy-tailed distributions are heavily right skewed, with a minority of large values in the head and a majority of small values in the tail, commonly characterized by a power law, a lognormal or an exponential function. For example, a country's population is often distributed in such a heavy-tailed manner, with a minority of people (e.g., 20 percent) in the countryside and the vast majority (e.g., 80 percent) in urban areas. This heavy-tailed distribution is also called scaling, hierarchy or scaling hierarchy. This new classification scheme partitions all of the data values around the mean into two parts and continues the process iteratively for the values (above the mean) in the head until the head part values are no longer heavy-tailed distributed. Thus, the number of classes and the class intervals are both naturally determined. We therefore claim that the new classification scheme is more natural than the natural breaks in finding the groupings or hierarchy for data with a heavy-tailed distribution. We demonstrate the advantages of the head/tail breaks method over Jenks' natural breaks in capturing the underlying hierarchy of the data.

## 1. Introduction

Human thinking often involves binary thinking or dualism, which divides things or phenomena into two opposing categories, such as urban/rural, high/low, long/short, and extraordinary/ordinary. Many societal and natural phenomena demonstrate, and can therefore be perceived as, dual parts consisting of an imbalanced contrast; one part is often distinguished as the background (the majority) and the other as the foreground (the minority). For example, in terms of occupied land area, there are far more rural areas than urban areas; 90 percent of the land in Europe is countryside, while only 10 percent is urban. On the other hand, far more people live in urban areas than rural areas; 80 percent of people in America and Europe are urban residents and only 20 percent live in the countryside. We also know that there are far more ordinary people (say, 80 percent) than extraordinary people (say, 20 percent); this is often characterized by the 80/20 principle, based on the observation made by the Italian economist Vilfredo Pareto in 1906 that 80% of land in Italy was owned by 20% of the population. A histogram of the data values for these phenomena would reveal a right-skewed or heavy-tailed distribution. How to map the data with the heavy-tailed distribution?

There are a range of data classification methods for statistical mapping, such as equal steps, quantiles, geometric progressions, standard deviation and natural breaks (Coulson 1987, Evans 1977). The natural breaks method is used to classify data values into different classes according to the breaks or gaps that naturally exist in the data. This is done by plotting the frequency of the data values and then identifying the breaks in the data. This idea of relying on the data frequency to identify class intervals has existed for a long time (Alexander and Zahorchak 1943). To illustrate the method, we examine a statistical dataset of rural population densities utilized by George Jenks in an early study (Jenks 1963). Figure 1 illustrates the original data with 1a and 1b respectively depicting the spatial distribution and the statistical distribution. With a few exceptions, most of the data values are centered around 5.5 and lie between 1.6 and 15. The exceptions are centered on 18.0, 28.0, 35.0, and 103.4. There are some gaps between the clustered data values, with one of the gaps (between 35.0 and 103.4) being very large. These gaps constitute the class intervals of the natural breaks. Based on the natural breaks, Jenks (1967) developed an optimization method to minimize within-class variance while maximize between-class variance; this is similar to the concept adopted by Fisher (1958) for data classification. This optimization method is also known as the goodness-of-variance-fit (GVF) method and is achieved through iterative computing of the GVF; that is, moving one data value from the class with the largest deviations from the array mean to the class with the lowest deviations from the array mean until the sum of the within-class deviations reaches a minimal value. The resulting classification is called Jenks' natural breaks and is widely used in statistical mapping.

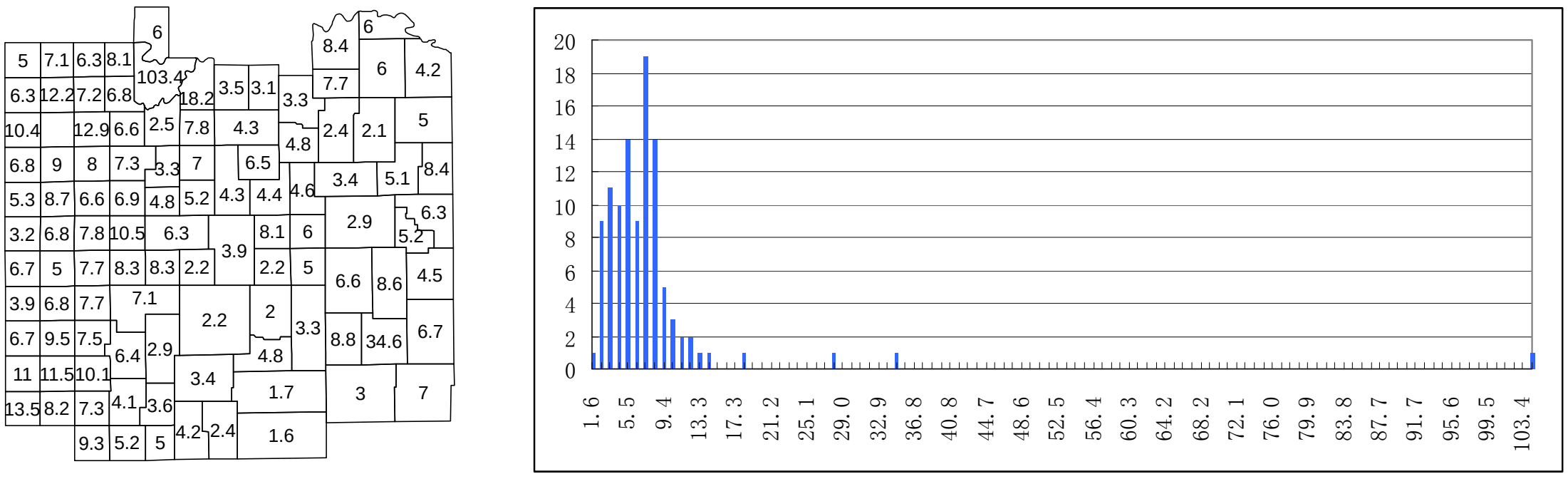

(a) (b)

Figure 1: (Color online) Rural population densities: (a) geographic distribution of the data, and (b) statistical distribution of the data
(Note: The data concerns the population densities for 105 minor civil divisions in eight central Kansas counties)

The above discussion shows that data classification involves two issues: the number of classes and the class intervals. This paper proposes a classification scheme that naturally determines the number of classes and class intervals based on human binary thinking and applies to data that follow a heavy-tailed distribution. The heavy-tailed distribution is commonly found in many societal and natural phenomena, including geographical systems. Small events are far more common in geographic spaces than large events (Jiang 2010), particularly in urban, architectural environments (Salingaros and West 1999, Jiang 2009). For example, there are far more small cities than large ones (Zipf 1949); far more short streets than long ones (Jiang 2009); far more small city blocks than large ones, and far more low-density areas than high-density ones (Jiang and Liu 2011). This is the pattern that underlies geographic spaces and which should be reflected in the map.

The remainder of this paper is organized as follows. The next section introduces a set of heavy-tailed distributions and elaborates on how they differ from the normal distribution. After that, we propose the head/tail breaks classification scheme and argue that it reflects the underlying hierarchy of heavy-tailed distributions. We then present two case studies mapping U.S. cities and Swedish streets to further illustrate how the new classification scheme reflects the underlying patterns of data. We further discuss the importance of data classification in statistical mapping and several advantages of the head/tail breaks before concluding the paper.

## 2. Heavy-tailed distributions

Existing classification methods are dominated by a Gaussian way of thinking; they focus on high-frequency events, but consider low-frequency events to be separated from high-frequency ones. There are large gaps or breaks (shown between the blue bars in Figure 1b) between high-frequency and low-frequency data values, as well as between different levels of low-frequency values. These gaps and breaks constitute the foundation for the natural breaks classification. This paper reverses this way of thinking by concentrating on low-frequency values. The lowest frequency value is ranked number 1, the second-lowest frequency value is ranked number 2, and so on. The ranking is plotted on the x-axis and the corresponding values on the y-axis. This plot, known as the rank-size distribution in general and initially used for city sizes and word frequencies (Zipf 1949), is a typical heavy-tailed distribution (Figure 2)

We focused on low-frequency events because they tend to have high impact, which makes them more important than high-frequency events. Low-frequency events contain far more information than high-frequency events because of their improbable nature. For example, news headlines are low-frequency events that occur rarely and are therefore newsworthy; they convey more information than high-frequency events. Nature, society, and our daily lives are full of rare and extreme events, which are termed "black swan events" (Taleb 2007). This line of thinking provides a good reason to reverse our thinking by focusing on low-frequency events.

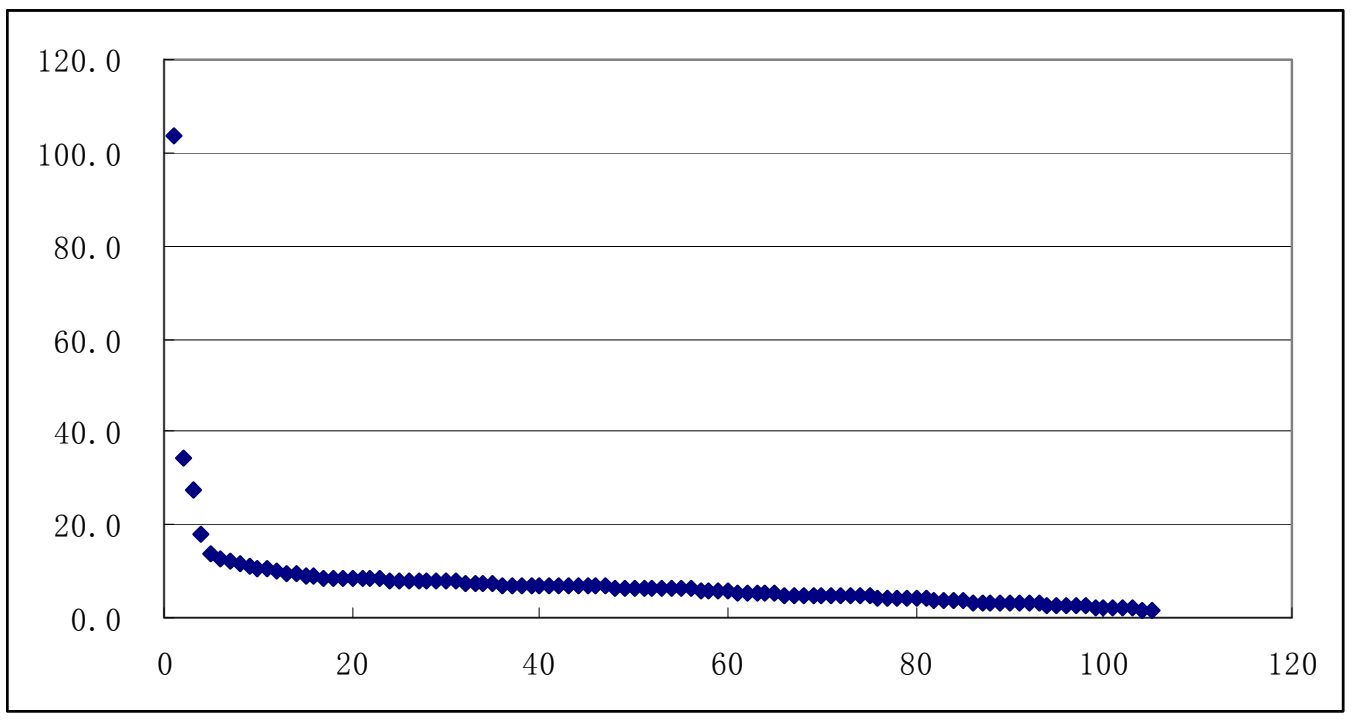

Figure 2: A typical heavy-tailed distribution based on the rural population density data

Unlike a normal distribution, which is bell-shaped and has two thin tails that reach the x-axis rapidly, a heavy-tailed distribution theoretically has a long tail skewed toward the right that approaches but never touches the x-axis. The entire distribution line or the long tail consists of the head and the tail, split around an average of the data values. Importantly, the head is composed of a minority of the large data values, whereas the tail is composed of a majority of the small data values. This imbalance between the head and the tail (or the minority and the majority) has been formulated as the head/tail division rule; that is, "*given a variable X, if its values x follow a heavy tailed distribution, then the mean (m) of the values can divide all the values into two parts: a high percentage in the tail, and a low percentage in the head*" (Jiang and Liu 2011). Figure 3 illustrates a heavy-tailed distribution in which the head and tail is divided around the mean value m, with only 10 percent data values in the head and 90 percent data values in the tail. The use of the mean or average to partition the data values into two imbalanced parts is in line with binary thinking of the human mind.

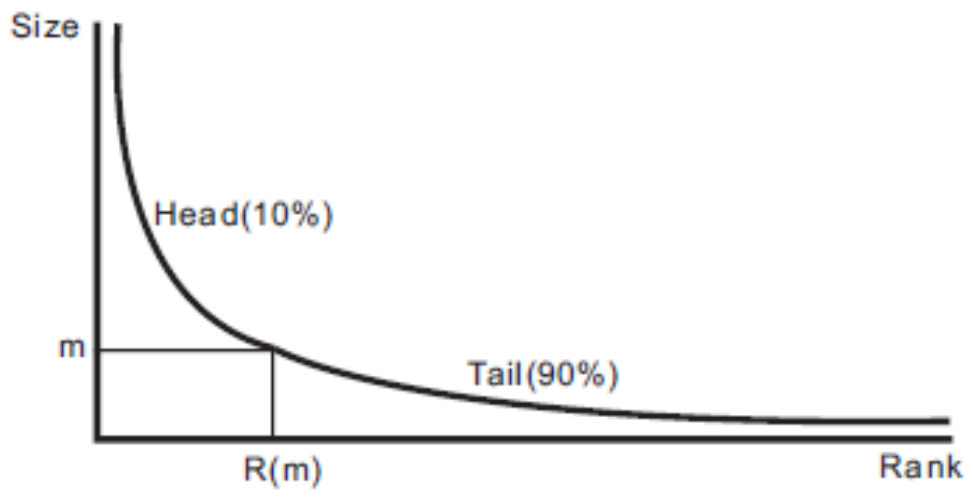

Figure 3: Illustration of the head/tail division rule
(Note: A rank-size distribution with the x-axis as the ranking and the y-axis as the corresponding ranked data values. The arithmetic mean is m and the corresponding ranking is R(m). The head and the tail contain imbalanced amounts of data values.)

In general, a heavy-tailed distribution can also be observed when assessing the probability density function (PDF) of the data values. It should be noted that low-frequency events are in the head in rank-size distributions, but in the tail in PDFs. In this paper, to avoid confusion, the terms head and tail refer to the rank-size distribution rather than the PDF. Given a variable of interest x and the probability of occurrence of the variable value y, the most typical heavy-tailed distribution is a power law, expressed as

$$y = x^{-\alpha}, \qquad [1]$$

where α is termed the power law exponent.

Taking the logarithm of both sides in Equation [1], we obtain

$$\ln y = -\alpha \ln x. \qquad [2]$$

The distribution line is straight when graphed as a double-logarithm plot. This straight distribution line, which is the hallmark of a power law, is often used to detect the power law in double-logarithm plots, so called log-log plots. However, this method, which is essentially based on the least squares method, has been criticized for creating errors or bias when drawing conclusions regarding the existence of a power law distribution. Other methods based on the maximum likelihood have been proposed as being more appropriate for detecting a power law distribution. In this paper, we adopt the latter method in detecting a heavy tailed distribution. It should be noted that we keep mathematics at the simplest level; interested readers should refer to Clauset et al. (2009) and the references therein for more details.

A heavy-tailed distribution can also be characterized by the lognormal function, which has the format

$$y = \frac{1}{\sqrt{2\pi\sigma^2}\,x}\exp\left(\frac{-(\ln(x)-\mu)^2}{2\sigma^2}\right), \qquad [3]$$

where the parameters $\mu$ and $\sigma$ are the mean and standard deviation of ln(x), respectively. This is a normal distribution for ln(x), but the variable x itself is heavy-tailed distributed; that is, lognormal.

In the literature, particularly in the statistical physics literature, the tail of a heavy-tailed distribution is considered to be heavier or more long-tailed than that of an exponential distribution. The "heavier" or "more long-tailed" is what we referred to early for the tail approaching but never touching the x-axis. Therefore, the exponential distribution is excluded from the class of heavy-tailed distributions with a good reason. Physicists believe that the mechanisms underlying events that follow the "strictly defined" heavy-tailed distribution are fundamentally different from those that follow an exponential distribution.

In the simplest style, an exponential distribution is expressed as

$$y = e^{x}. \qquad [4]$$

Taking the logarithm of both sides in Equation [4], we obtain

$$\ln y = x. \qquad [5]$$

Equation [5] indicates that a linear relationship exists between x and the logarithm of y. This is different from the power law distribution, for which there is a linear relationship between the logarithm of x and logarithm of y. However, this paper loosely considers the exponential distribution to be a part of the heavy-tailed distribution family, for two reasons: (1) each of the three functions illustrated above are right-skewed, and (2) it is extremely difficult sometimes to differentiate among the three distributions with real-world datasets. Instead of following any of the three distributions exactly, real world datasets often follow degenerate heavy-tailed distributions, such as a power law with an exponential cutoff or a stretched exponential distribution (Clauset et al. 2009, and references therein). In the present study, all heavy-tailed distributions were considered to be a good approximation of a power law function.

## 3. A novel data classification scheme: head/tail breaks

This paper proposes a data classification scheme in which the number of classes and the class intervals are both naturally determined by the very property of the heavy-tailed distributions of the data. As discussed above, a heavy-tailed distribution consists of a head and a tail, each of which contains an unbalanced proportion of data values. This imbalance between the head and tail, or between many small values and a few large values, can be expressed as “far more small things than large things.” Perceptually, the tail constitutes a background upon which the head is distinguished. Interestingly, this imbalance recurs in the head; the values in the head can be divided again around an arithmetic mean into many small values and a few large values. In other words, the values demonstrate a hierarchical structure. This inherent hierarchy constitutes the foundation of the proposed classification scheme, which is termed here as “head/tail breaks”.

The head/tail breaks method groups the data values into two parts around the arithmetic mean and continues the partitioning for values above the mean iteratively until the head part values are no longer heavy-tailed distributed. Therefore, the number of classes and the class intervals are both naturally determined by the data, or the data’s very property of heavy-tailed distribution. In this way, the head/tail breaks classification method differs fundamentally from the nested means classification method, which iteratively partitions the values in both the head and the tail (Scripter 1970). The nested means starts by partitioning the data values into two classes around the mean: those above the mean and those below the mean. This process continues for each of the two classes and using the mean for each class will result in four classes. This process can continue to get eight classes, 16 classes, and so on. In fact, there is no need to further classify or partition the values in the tail or below the mean because the variation in the tail is small enough.

This classification scheme is illustrated by applying it to the above-mentioned rural population density data. As noted, the data are heavy-tailed distributed. The first arithmetic mean is 7.6, which partitions all the data into two groups: those above the mean or in the head, and those below the mean or in the tail. The values in the head are also heavy-tailed distributed. The second mean, for the values in the head, is 14.0. Four values exceed 14.0; namely, 18.2, 27.4, 34.6, and 103.4. The third mean, for these four values, is 45.9, which can be used to partition the four values into two groups, in which the value 103.4 constitutes one group. Eventually, the classification scheme ends up with four classes, as shown in Figure 4a, which is put in comparison with the natural breaks classification (Figure 4b).

It is worth elaborating on the pattern illustrated by the head/tail breaks. The first class ranges from the minimum value (1.6) to the first mean (7.6), and constitutes the background for the second class. The second class ranges from the first mean (7.6) to the second mean (14.0) and constitutes the background for the third class. The third class ranges from the second mean (14.0) to the third mean (45.0) and constitutes the background for the fourth class. The fourth class ranges from the third mean (45.0) to the maximum value (103.4) and constitutes the foreground. The lowest (or first) class is the background and the highest (or fourth) class is the foreground. All of the other middle classes serve as both a background for the immediate upper class and a foreground for the immediate lower class. Importantly, both the backgrounds and the foregrounds, which correspond to the tail and the head, create an imbalanced contrast; that is, a small foreground versus a vast large background. This is exactly the underlying pattern for data with a heavy-tailed distribution. This pattern can also be called scaling or hierarchy, which should be retained in the map.

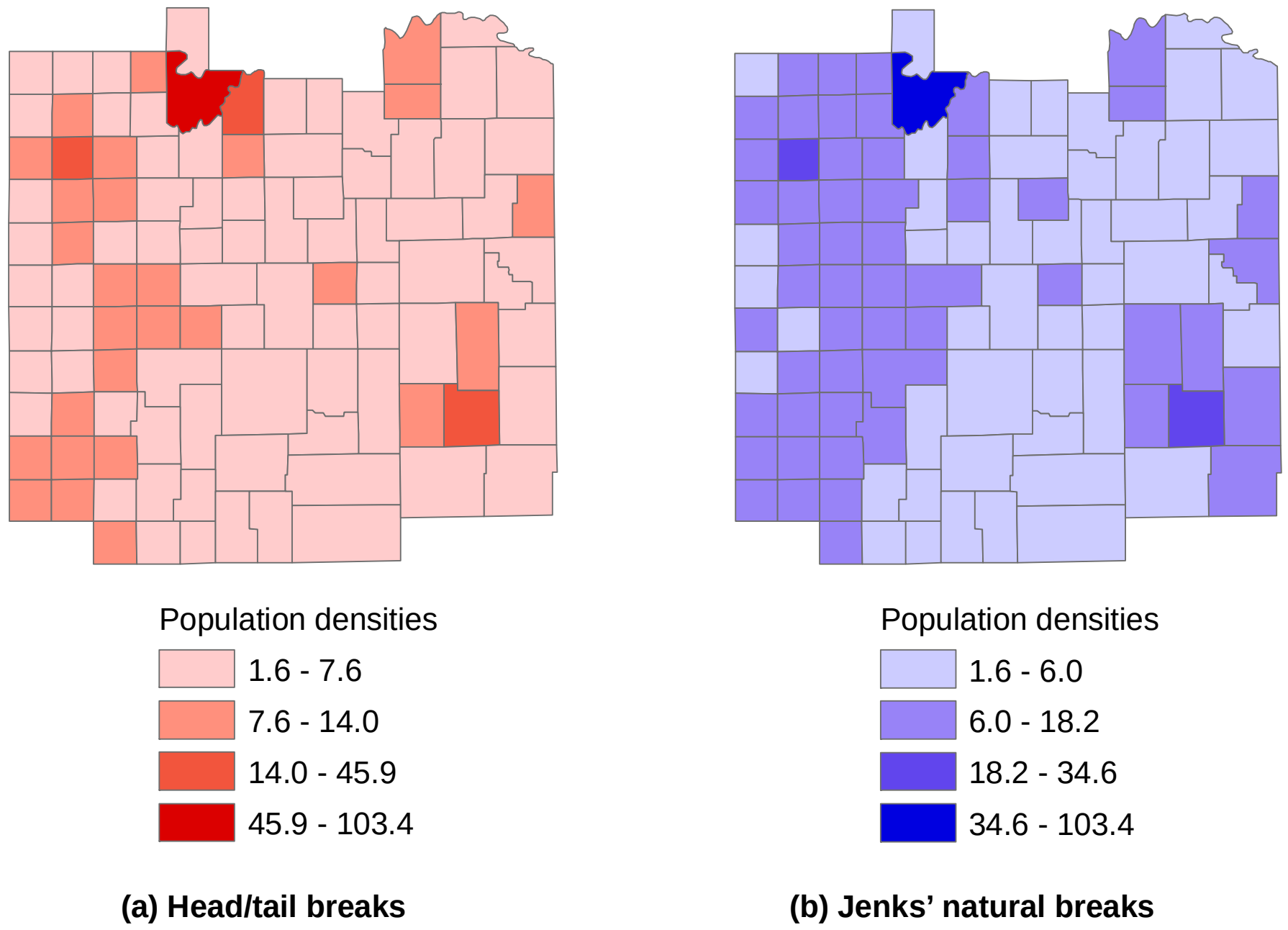

Figure 4: (Color online) Head/tail breaks (a) compared with Jenks' natural breaks and (b) using the rural population density data

A good classification method should largely reflect the pattern that underlies the data. In other words, any classification method that distorts the underlying pattern of data is considered a poor classification. The pattern underlying heavy-tailed distributed data is scaling or hierarchy, which can be expressed as "far more smaller things than larger ones". The notion of 'scaling of geographic space' refers to the hierarchy that is characterized by the heavy-tailed distribution (Jiang 2010). The pattern illustrated by Jenks' natural breaks (Figure 4b) clearly lacks this scaling property; for example, the two lowest classes occupy approximately same percentage of areas, so no hierarchy is shown. It should be noted that the expression "far more smaller things than larger ones" can sometimes be stated the other way around. For example, in a city center where almost all buildings are skyscrapers, it could be stated as "far more larger things than smaller ones." In this case, the minority of smaller values constitute the head and the majority of the larger values constitute the long tail.

## 4. Case studies: head/tail breaks in comparison with Jenks' natural breaks

We conducted two case studies to further illustrate the advantages of the head/tail breaks scheme in capturing the underlying hierarchy of the data. The two case studies contain considerably more data values than the small example shown above. This is because the heavy-tailed distributions usually appear in systems that are large in terms of time or space, which implies that large systems have evolved over a long time period or developed in a large spatial setting. It is often the case that the larger the systems, the more obvious or striking the heavy-tailed distributions. Another reason is the perspective from which we observe the world. For example, a street network is not heavy-tailed distributed in terms of the degree of junction connectivity, but it is heavy-tailed in terms of the degree of the connectivity of individual streets (Jiang et al. 2008). The correct perspective and scope must be chosen in order to observe the scaling of geographic space.

The first case study concerns the statistical mapping of U.S. cities. According to data from the 2000 U.S. census (e.g., National Weather Service 2011), there are 42,934 cities in the United States. For the purposes of this study, 5242 cities with populations greater than 8000 were selected for mapping. The selected city sizes are ranked by population; the smallest city has 8,000 people and the largest city has 7,322,564. The total population in the 5242 cities is 284,916,464. The city sizes do not appear to be normally distributed; there are two reasons for this. Firstly, both the smallest size and the largest size are far from the average size of the city, which is 54,353 (284,916,464/5242 = 54,353). Secondly, the ratio of the largest size to the smallest size is very large: 7,322,564/8000 = 915. Further examination of the city sizes demonstrates that they are indeed power-law distributed. Figure 5 illustrates the log-log plot, which indicates that the power law fits across two decades of the data values.

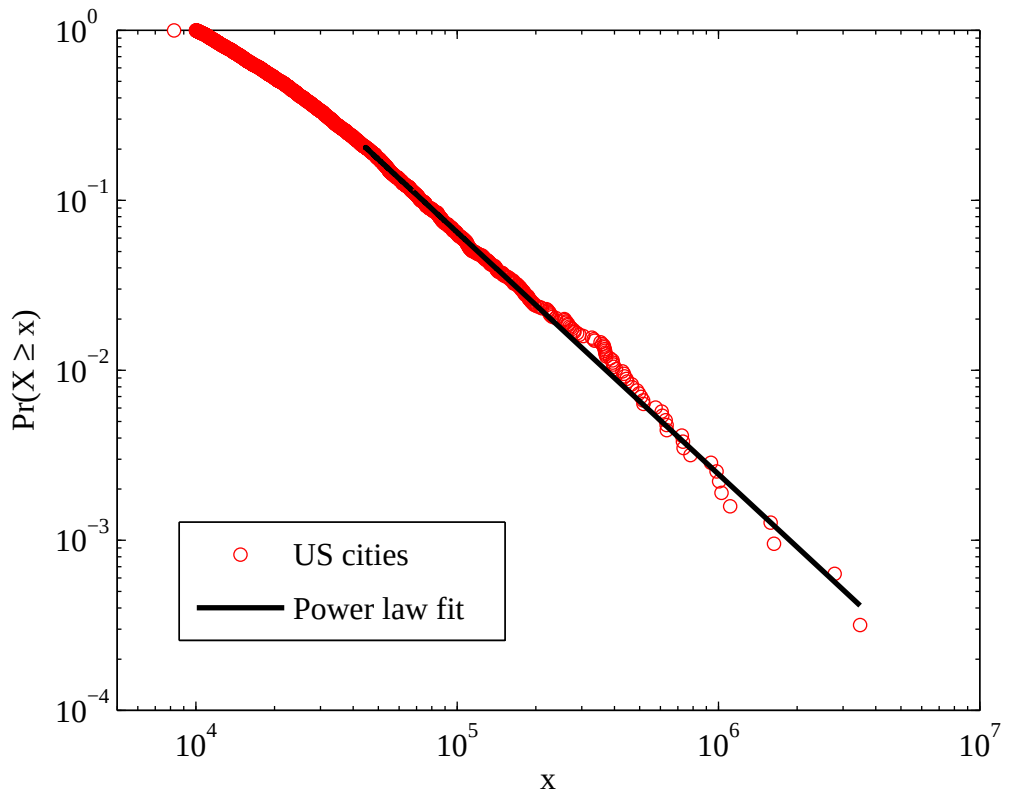

Figure 5: (Color online) Power law distribution of U.S. city populations

We further examine the hierarchy of the data values. The first arithmetic mean (54,353) groups all the cities into two categories: those above the mean and those below the mean. There are 744 cities above the mean, which accounts for 14 percent (a minority) of all cities. The second mean on these 744 cities is 268,370. This value can be used to further partition the 744 cities into two categories. Of the 744 cities above the mean, only 121 (16 percent) have populations larger than the average of 268,370 (again, a minority). The third mean, 1,111,291, derived from the 121 cities, partitions the 121 cities into two groups, the result of which is shown in Table 1. The 18 cities in the last head are no longer heavy-tailed distributed, which means it is necessary to stop the partitioning. Eventually, the three partitions lead to the creation of four classes or hierarchical levels. The four class intervals are determined by the smallest size (8000), the largest size (7,322,564), and the three means: [8000, 54,353], [54,354, 268, 370], [268,371, 1,111,291], [1,111,292, 7,322,564].

Table 1: Three partitions leading to four hierarchical levels of U.S. cities

| # of cities | # in head | % in head | # in tail | mean |
|---|---|---|---|---|
| 5,242 | 744 | 14% | 4,498 | 54,353 |
| 744 | 121 | 16% | 623 | 268,370 |
| 121 | 18 | 15% | 103 | 1,111,291 |

The four classes are mapped using graduated circles and compared with classes generated using Jenks' natural breaks scheme (Figure 6). The map illustrated by the head/tail breaks reflects the underlying pattern better than the map generated using Jenks' natural breaks. The pattern based on Jenks' natural breaks look very flat (not much variance), whereas the map based on the head/tail breaks is full of changes or surprising changes, both locally and globally. This scaling pattern is more obvious when it is assessed with Table 1. In view of this, the map based on the head/tail breaks scheme reflects the hierarchy of the data values much better than the one based on Jenks' natural breaks.

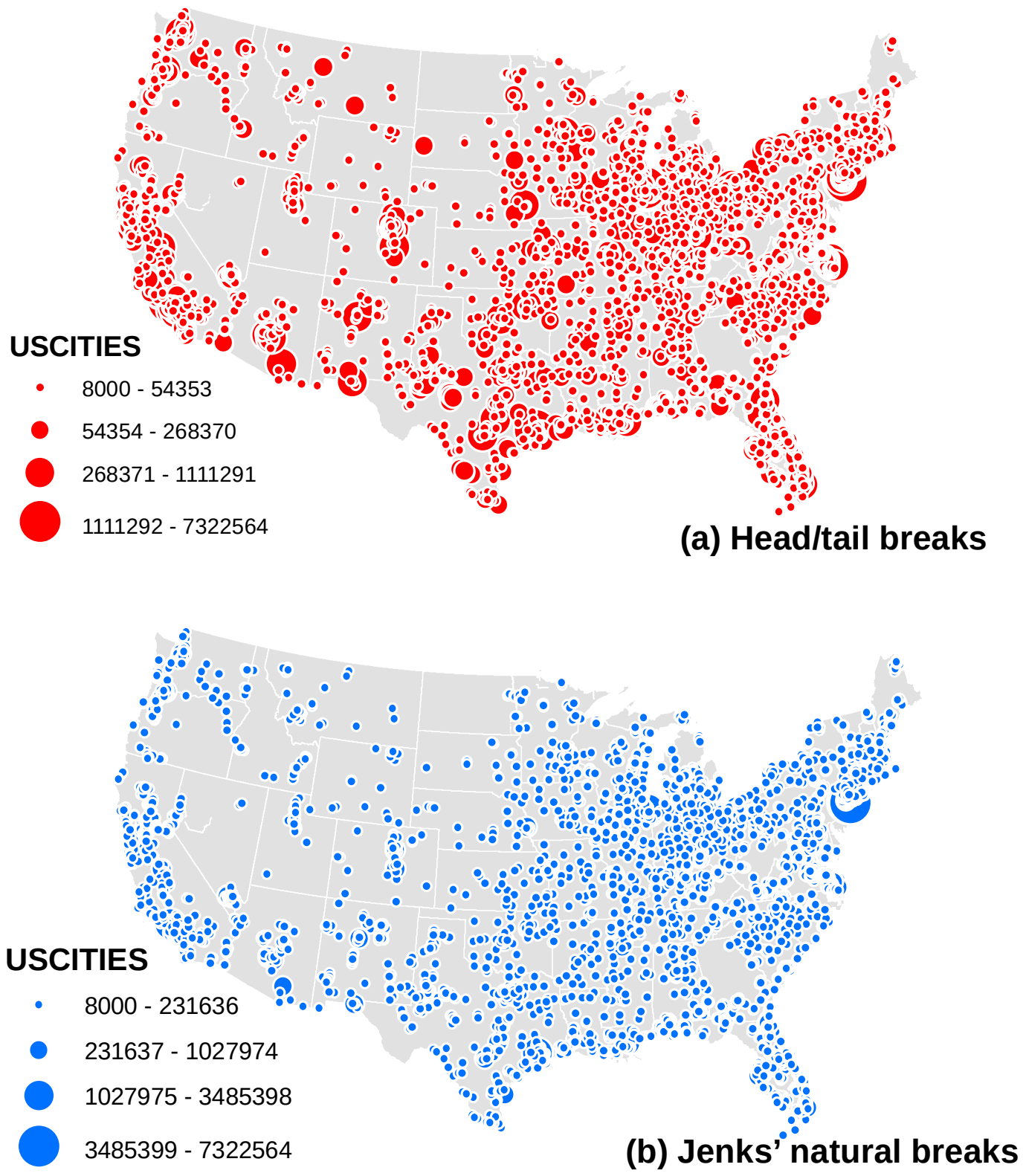

Figure 6: (Color online) Classification of U.S. cities based on (a) head/tail breaks and (b) Jenks' natural breaks

The second case study maps the underlying hierarchy of 160,000 Swedish streets. The streets are either named streets with unique names, or natural streets that constitute good continuity in perception (Jiang and Claramunt 2004, Jiang et al. 2008). The degree of connectivity of individual streets was mapped; that is, the number of other streets that connect or intersect each individual street. The degree of connectivity ranges from 1 to 1040. Therefore, the ratio of the most connected to the least connected streets is 1040/1 = 1040, which is a clear indicator of a heavy-tailed distribution. A less pronounced fit to a power law distribution was uncovered as illustrated in Figure 7; it is definitely a heavy tailed distribution. Due to the distribution, the data values can be partitioned nine times around different means. Table 2 shows the result of the classification, where the head always accounts for a minority of streets. The nine partitions lead to 10 classes or hierarchical levels, which are mapped using a spectral color legend as shown in Figure 8.

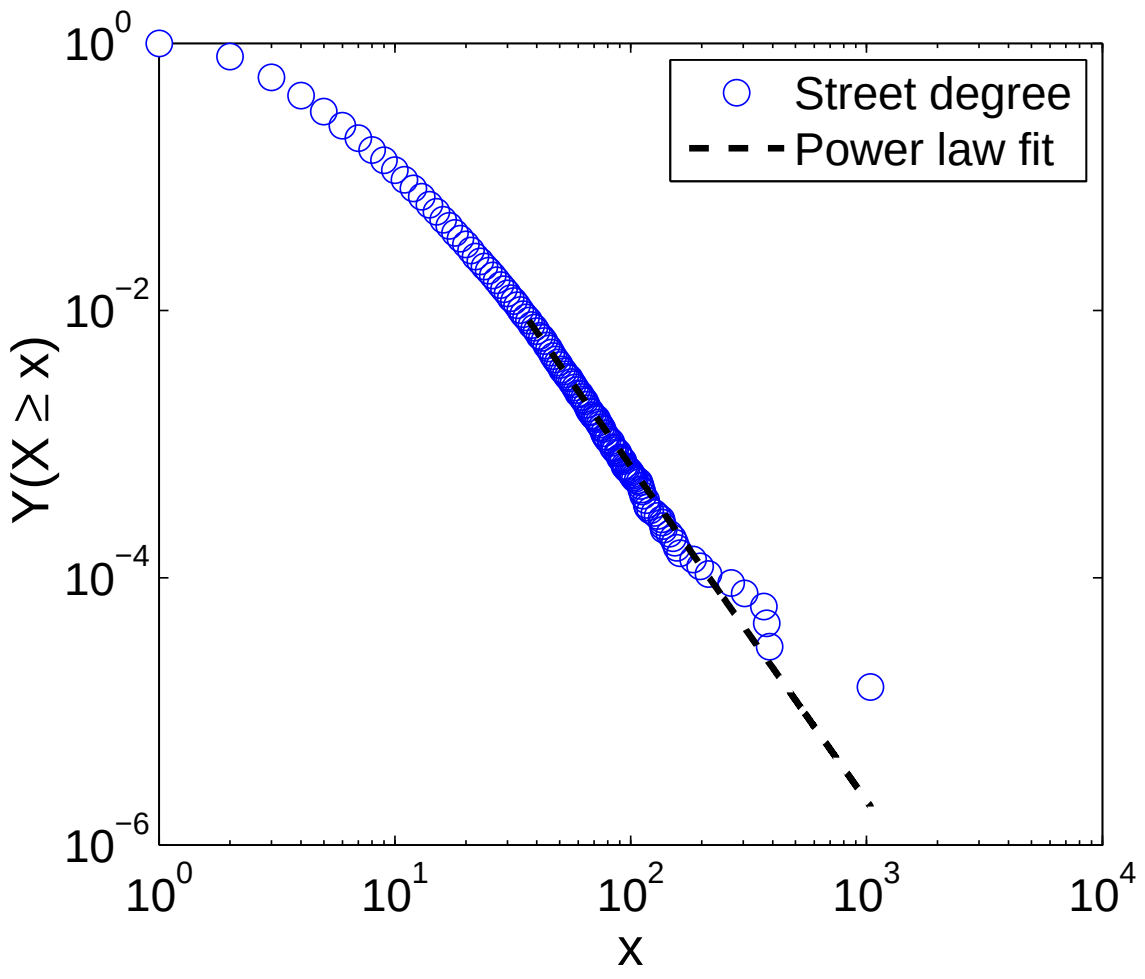

Figure 7: (Color online) Heavy tailed distribution for the connectivity of Swedish streets

Table 2: Nine partitions leading to 10 classes of Swedish streets

| # of streets | # in head | % in head | # in tail | mean |
|---|---|---|---|---|
| 166,479 | 47,021 | 28% | 119,458 | 3.6 |
| 47,021 | 14,382 | 31% | 32,639 | 8.0 |
| 14,382 | 4,235 | 29% | 10,147 | 14.8 |
| 4,235 | 1,294 | 31% | 2,941 | 26.0 |
| 1,294 | 363 | 28% | 931 | 43.2 |
| 363 | 101 | 28% | 262 | 71.7 |
| 101 | 21 | 21% | 80 | 119.8 |
| 21 | 6 | 29% | 15 | 239.6 |
| 6 | 1 | 17% | 5 | 457.4 |

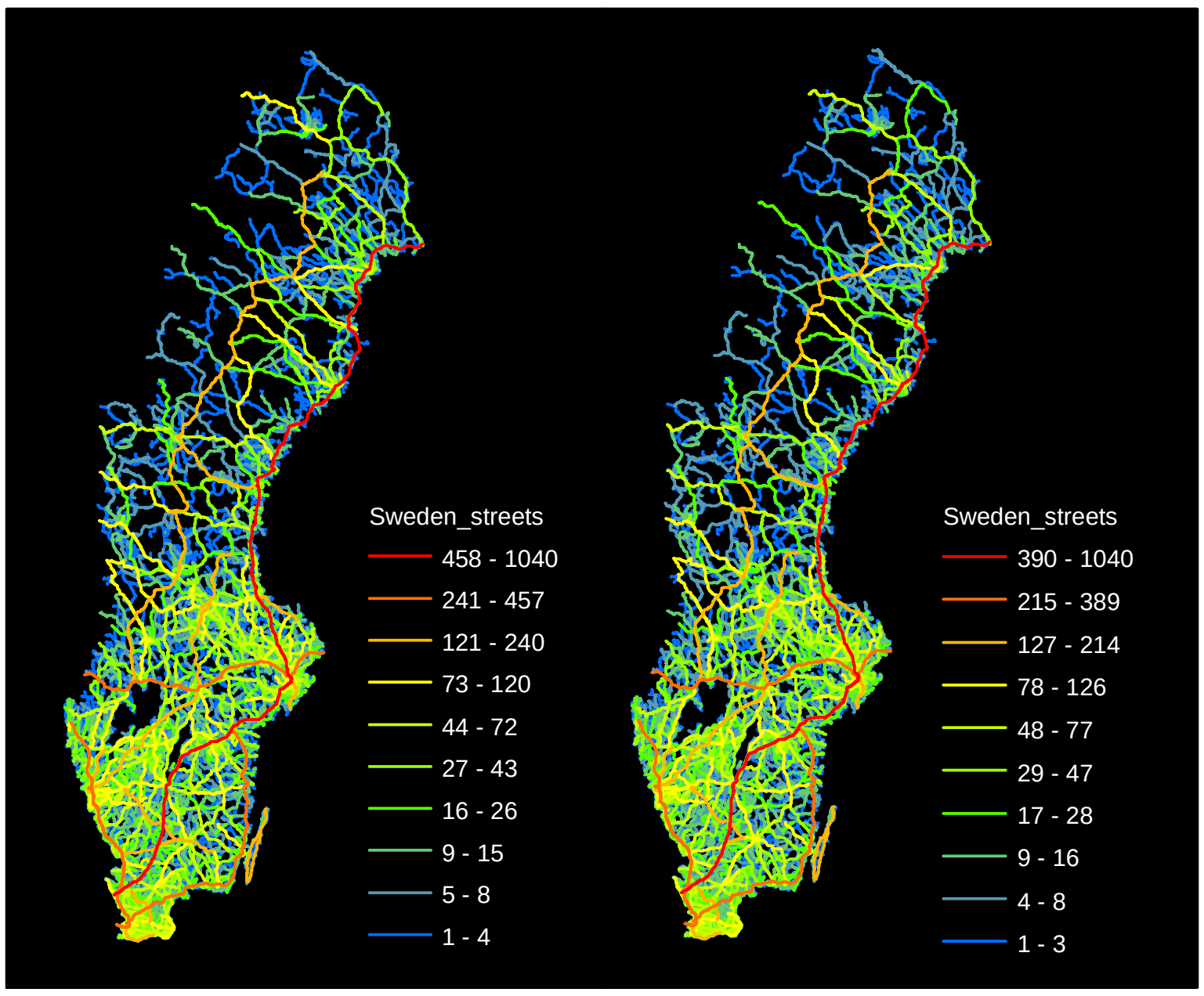

**(a) Head/tail breaks** **(b) Jenks' natural breaks**

Figure 8: (Color online) Classification of Swedish streets in terms of street connectivity based on (a) head/tail breaks and (b) Jenks' natural breaks

It is difficult to determine which of the two maps in Figure 8 is better solely on the basis of the visual effect because the high density of streets causes overlap. Most blue lines in the two maps are covered by lines in warm colors, such as red, orange, and yellow, which makes it difficult to visually assess the difference between the two maps. However, in terms of capturing the underlying hierarchy of the data values, the left pattern is indeed superior to the right one. This point can only be seen or be assessed from Table 2, which is associated with Figure 8. The right pattern does not adopt the inherent hierarchical levels; instead, it addresses the gaps that naturally exist in the data. However, the head/tail breaks scheme captures and reflects the underlying hierarchy more precisely than the natural breaks scheme.

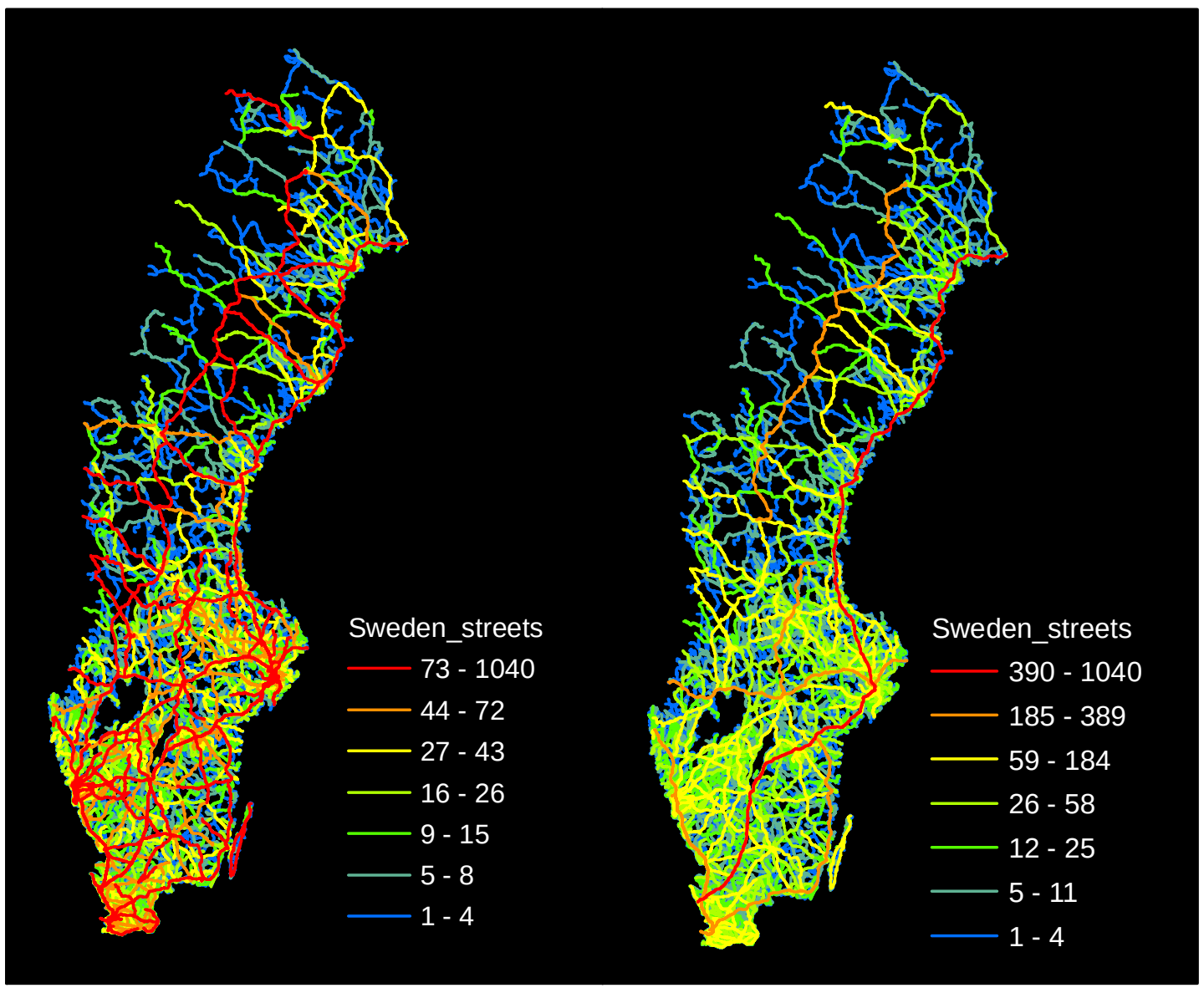

**(a) Head/tail breaks** **(b) Jenks' natural breaks**

Figure 9: (Color online) Reduction of the 10 classes into seven classes using (a) head/tail breaks and (b) Jenks' natural breaks

The second case study has 10 classes, which is slightly more than the human mind can hold in the memory simultaneously (Miller 1956). In this case, the number of classes can be reduced to seven (for example) by merging the top four classes into a single class. It is important to note that the merging does not affect the fundamental hierarchy. When this is performed, the two maps (based on the head/tail breaks and Jenks' natural breaks schemes) look quite different visually or in terms of the coloring (Figure 9). This is probably an advantage of the head/tail breaks compared to the natural breaks. Every time the number of classes changes, the natural breaks must re-compute all the class intervals by minimizing within-classes variance and maximizing between-classes variance. There is no need for this to happen with the head/tail breaks in order to maintain the derived hierarchy or classes. The upper classes are simply merged into one and all of the lower class intervals remain unchanged.

In perceiving a map such as Figures 9a or 8a, one should not be influenced by the visual effect, which incorrectly suggests that there are "far more long streets than short ones." In fact, there are far more short streets (which are, unfortunately, overlapped by long ones on the top) than long ones. In view of this and with respect to Table 2, the pattern on the left represents the hierarchy of the data, whereas the pattern on the right indicates the breaks in the data.

The two case studies have shown how the head/tail breaks classification scheme captures the inherent hierarchy of the data. Interestingly, a lower class constitutes a background in which the upper classes are distinguished. In other words, an upper class constitutes a foreground for the immediate lower class. Importantly, the background and the foreground are in an imbalanced contrast. This is in line with the binary thinking of the human mind. Therefore, this paper argues that the binary thinking-based classification is more natural than the natural breaks for data with a heavy-tailed distribution.

## 5. Discussion

Data classification is a fundamental issue in statistical mapping, and it is well known that different classification methods can lead to very different visual patterns, sometimes in a distorted or biased manner (e.g., MacEachren 1994, Monmonier 1996, Brewer and Pickle 2002). Because of the difficulty of classification, continuous classes or no-class-intervals maps have been suggested (Tobler 1973). Classification is an important step toward effective mapping, as human perceptual limits require effective classification in order to successfully map underlying patterns. One such visual pattern is the scaling pattern studied in this paper, which is pervasive in geographic space; that is, “far more small things than large ones.” This pattern is reflected in a great deal of geographic data. Mapping of geographic data should reflect the pattern of geographic space and a good classification scheme must capture the underlying pattern of the data distribution.

Data classification is, to a large extent, a process of generalization by differentiating between vital and trivial things. In fact, human thinking also generalizes when receiving information from the real world. For example, news headlines generally deserve more attention than trivial matters. This paper introduces a new way of considering data classification by concentrating on low-frequency things because these tend to be vital, important, and high impact. The head/tail breaks classification is driven by this new way of thinking; the resulting upper classes capture the important and vital things in the head, whereas trivial things are relegated to the tail as lower classes.

The head/tail breaks demonstrates several advantages compared with the natural breaks and other classification schemes. Firstly, the head/tail breaks classification scheme captures the image or hierarchy that appears in reality (far more smaller things than larger things). Secondly, the number of classes and the class intervals are naturally determined based on the hierarchy embedded in the data. The third advantage is that the number of classes matches well the human memory limit, which is approximately seven (Miller 1956). This third advantage contrasts strongly with the nested-means classification scheme, which creates too many classes because of the binary partitioning of the head and the tail and, and more critically, generates unnecessary and redundant breaks for the data values in the tail.

Some conventional classification methods pay attention to the scaling pattern or heavy-tailed distribution of data. Fundamentally, for data with a heavy-tailed distribution, class intervals cannot be imposed in a linear manner (for example, as equal steps), but rather in a nonlinear style. This nonlinear property is behind the use of nonlinear class interval methods, such as arithmetic and geometric progression methods. The widths of the class intervals increase in size at an arithmetic (or additive) rate. If the first class is two units, then the second and third classes would be four and six units, respectively. For the geometric progression, the widths of the class intervals increase in size at a geometric (or multiplicative) rate. If the first class is two units, then the second and third classes would be four and eight units, respectively. However, both arithmetic and geometric progression methods are not commonly used because neither the number of the classes nor the class intervals can be simply determined. More importantly, the class intervals imposed by the geometric regressions do not capture the underlying hierarchy.

## 6. Conclusion

This paper has proposed a novel classification scheme, the head/tail breaks, for data that are heavy-tailed distributed or right skewed. For the head/tail breaks scheme, the number of classes and class intervals are both naturally determined based on inherent hierarchical levels of the data. The class intervals are iteratively derived from the arithmetic means until the data values in the head are no longer heavy-tailed distributed. The head/tail breaks scheme captures the hierarchy of the data and is based on binary thinking of the human mind. In this sense, this classification scheme is more natural than the natural breaks for data with a heavy-tailed distribution.

The head/tail breaks method demonstrates a number of advantages over the commonly used methods for classifying data with a heavy-tailed distribution. The most important of these is that the classification scheme captures the essence of the data distribution that is scaling or hierarchy. Another notable advantage is the scheme’s simplicity; the class intervals are iteratively derived from the arithmetic means (c.f. Appendix). In this way, both the number of classes and the class intervals are naturally determined. If there are too many classes (for example, more than seven), the upper classes can simply be merged into a single class without affecting the lower classes. Although the advantages of the head/tail breaks method are clear, it remains to be seen how humans will respond to the new classification scheme compared to traditional ones.

## Acknowledgments

The author would like to thank the three anonymous referees and the editor Dr. Barney Warf for their useful comments that led to many improvements. Any shortcomings or inadequacies remain the responsibility of the author alone.

**Appendix: A short tutorial on how to perform the head/tail breaks classification in ArcMap**

Assuming that you have the USCITIES shapefile added into ArcMap as a map layer, open the attribute table associated with the shapefile. Find the field named “POP_1990”, right-click the column, and choose “Statistics…”. This will cause a pop-up window dialog that is similar to Figure 1A.

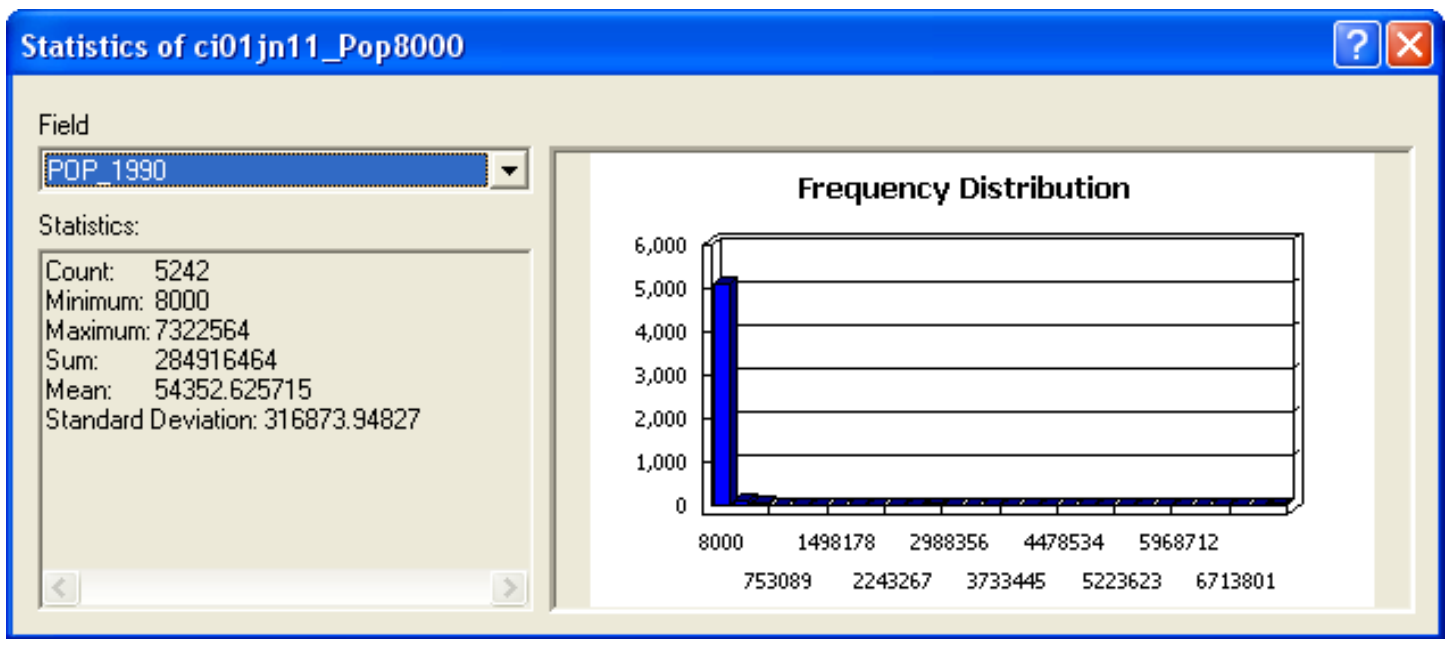

Figure A1: (Color online) Pop-up window for statistics

On the left-hand side of Figure A1, it can be seen that the total number of cities is 5242, the lowest size 8000, and the largest size 7,322,564. The frequency distribution to the right side looks L-shaped, which means it is a heavy-tailed distribution. Copy the first mean (54,353) and go to menu Selection > Select By Attributes. A window will appear that looks like Figure A2.

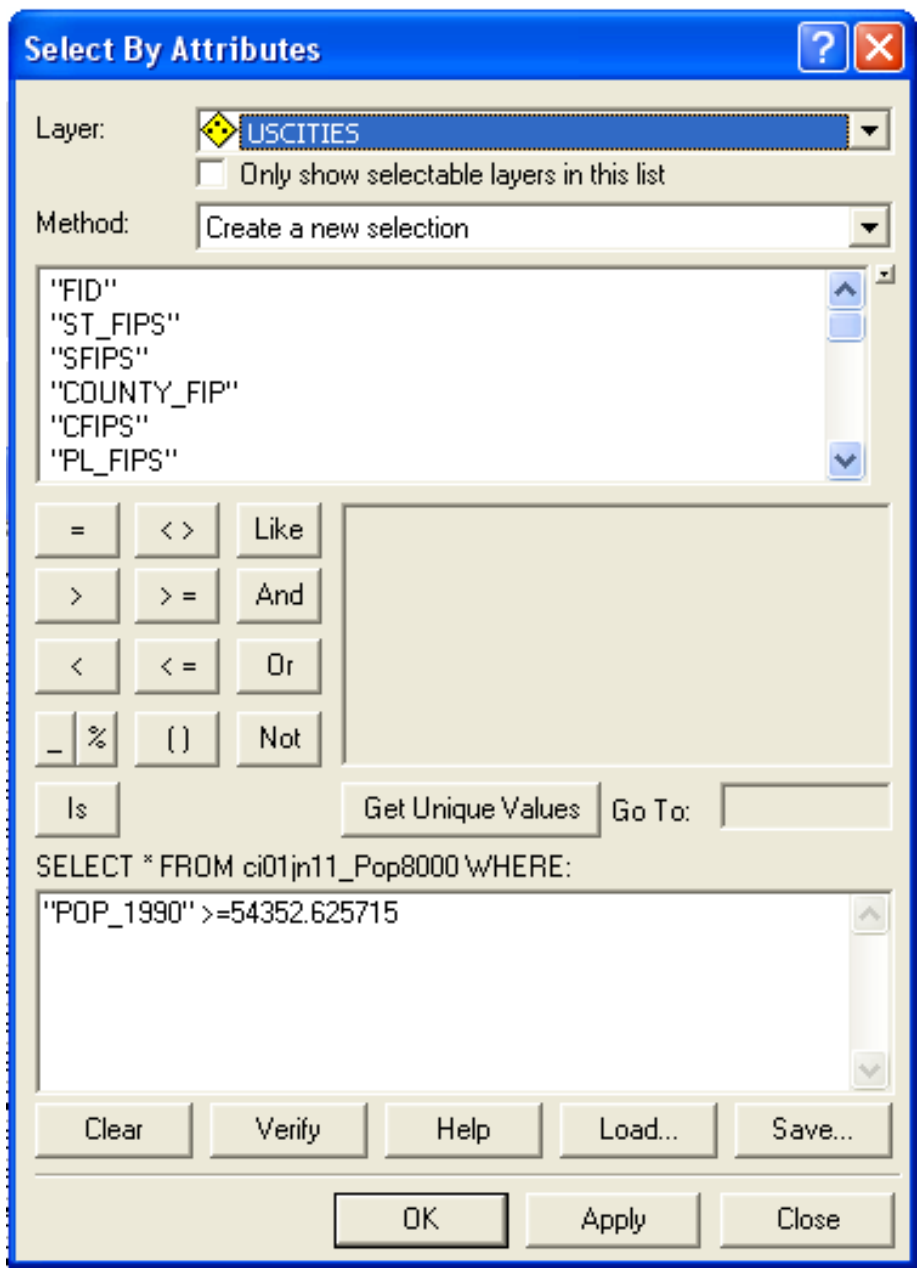

Figure A2: (Color online) Selection by attributes window

Choose the correct layer (e.g., USCITIES) and the correct field (e.g., POP_1990) for the selection “POP_1990” >= 54352.625715, as shown in Figure A2. This is how we selected larger cities above the first mean ($m_1$) or in the first head. Continuing this process will obtain the second and third means ($m_2$, and $m_3$). Next, we can conduct a manual classification by imposing the class intervals, [8000, 54,353], [54,354, 268, 370], [268,371, 1,111,291], [1,111,292, 7,322,564] in order to perform the classification. In general, the class intervals are determined by the lowest size (min), largest size (max), and the individual means (e.g., $m_1$, $m_2$, and $m_3$), i.e., [min, $m_1$], [$m_1$+1, $m_2$], [$m_2$+1, $m_3$], [$m_3$+1, max].